\documentclass[11pt]{article}
\usepackage{graphicx}
\begin{document}
\begin{center}
{\bf Low-lying isovector monopole resonances\footnote{first printed 
as preprint FT-353-(1989)/April, by the Institute of Physics and
Nuclear Engineering, Bucharest, Romania.}} \\[.5cm]
M. Grigorescu \\[1cm]
\end{center}
\noindent
$\underline{~~~~~~~~~~~~~~~~~~~~~~~~~~~~~~~~~~~~~~~~~~~~~~~~~~~~~~~~
~~~~~~~~~~~~~~~~~~~~~~~~~~~~~~~~~~~~~~~~~~}$ \\[.1cm]
{\bf Abstract}:
{\small Considering the equality of the proton and neutron Fermi levels as an
indication of a phenomenological interaction between pairs of protons and
neutrons, low-energy isovector monopole resonances are proved to appear in
the superfluid nuclei. The required phenomenological interaction is presented
as an isospin symmetry-breaking mean field for the four particle interaction.} \\
$\underline{~~~~~~~~~~~~~~~~~~~~~~~~~~~~~~~~~~~~~~~~~~~~~~~~~~~~~~~~
~~~~~~~~~~~~~~~~~~~~~~~~~~~~~~~~~~~~~~~~~~}$ \\[.5cm]
{\it 1. Introduction} \\[.3cm] \indent
The isovector monopole resonances (IVMR) were predicted in the framework of the
nuclear hydrodynamic model \cite{1,2} as out-of-phase breathing oscillations
of the proton and neutron densities. These oscillations represent the isovector
counterpart to the isoscalar compression modes, and are generated by the
restoring force which appears from the symmetry energy. Experimentally, IVMR
were first observed in the charge exchange reactions $^{90}$Zr, $^{120}$Sn
$(\pi^-, \pi^0)$ at $T_{\pi^-}=165$ MeV \cite{3}, and were also confirmed
recently  in the $(n,p)$ reactions on $^{90}$Zr \cite{4}. Other experiments
\cite{5} have shown their occurrence for a wide range of nuclei, from
$^{40}$Ca to $^{208}$Pb, both normal and superfluid. \\ \indent
The quantum description of the superfluid systems shows a close connection
between the velocity potential $\chi$, ($\vec{v} = - \vec{\nabla} \chi$),
describing the flow and the gauge angle $\varphi$ from the BCS transformation to
quasiparticles \cite{2}: $\vec{\nabla} ( \chi - \varphi /2m)=0$, where $m$ is the nucleon
mass. For the monopole vibrations, this connection implies a radial
dependence of the proton and neutron gauge angles $\varphi_p$, $\varphi_n$.
In contrast to the space dependence, the time dependence of $\varphi_\tau$,
$\tau=p,n$ occurs even if no macroscopic flow is present ($\vec{\nabla}
\chi=0$). The time-derivative $\dot{\varphi}_\tau /2$ is just the Fermi energy
$\lambda_\tau$ \cite{2}, and a superfluid system performs in its ground state
a free gauge rotation with the angular velocity $\dot{\varphi}_\tau=2
\lambda_\tau$. As is known, the proton and neutron systems are not closed,
changing particles through $\beta$ decay up to the equalisation of the Fermi
energies $\lambda_p$, $\lambda_n$. This can be taken as an indication of a
phenomenological gauge-restoring interaction, tending to fix not only the space-
depending term, but the whole relative angle $\varphi_p-\varphi_n$ at a constant
value. Consequently, low-lying IVMR, appearing only in superfluid nuclei
are expected, as the isovector counterpart to the isoscalar uniform gauge
rotation. The phenomenological character of the gauge-restoring interaction
generating these low-lying modes is imposed by its non-commutation with the
particle-number operators $\hat{N}_p$, $\hat{N}_n$. So, it can be considered
only as a mean-field approximation for a microscopic number-conserving Hamiltonian.
\\ \indent
In this paper section 2 presents the phenomenological gauge interaction
as a mean field deriving from a microscopic four-particle interaction. Restoring
the isospin symmetry by the cranking method, a kinetic energy term
contributing to the symmetry energy is obtained. These connections between the
gauge-restoring interaction, the four-particle interaction, and the symmetry
energy are used at the end of section 2 to estimate the former's strength
$\delta_0$. Following the semiclassical treatment of the proton-neutron
quadrupole-quadrupole interaction at deformed nuclei \cite{6} in section 3 the
effects of the gauge interaction in superfluid nuclei are investigated. This
approach gives both the underlying classical picture and the quantised spectrum
of the Fermi levels oscillations, allowing for realistic estimates of their
energy. \\[.3cm]
{\it 2. The gauge-restoring interaction} \\[.3cm] \indent
The early attempts to construct an isoscalar four-particle interaction
were based on the algebraic properties of the pairing operators, of having
closed commutation relations with the isospin operators. Denoting by $(a,m)$,
$a \equiv (n,l,j)$, the shell-model states $(n,l,j,m)$, the proton-proton,
neutron-neutron and the proton-neutron pair creation operators $P_+$, $N_+$,
$R_+$, are:
\begin{eqnarray}
P_+ & = & \frac{1}{2} \sum_{a,m} s_{am} c^\dagger_{pam} c^\dagger_{pa-m}~~,~~
s_{am}=(-1)^{j-m} \\
N_+ & = & \frac{1}{2} \sum_{a,m} s_{am} c^\dagger_{nam} c^\dagger_{na-m} \\
R_+ & = & \frac{1}{2} \sum_{a,m} s_{am} c^\dagger_{nam} c^\dagger_{pa-m}~~.
\end{eqnarray}
If $P_-$, $N_-$, $R_-$ are their hermitian conjugates, then
\begin{equation}
[P_+,P_-]=2P_0~~,~~ [N_+,N_-]=2N_0~~,~~[R_+,R_-]=2R_0
\end{equation}
with
\begin{eqnarray}
P_0 & = & \frac{1}{2} \sum_{a,m}( c^\dagger_{pam} c_{pam}- \frac{1}{2}) \\
N_0 & = & \frac{1}{2} \sum_{a,m} ( c^\dagger_{nam} c_{nam}- \frac{1}{2}) \\
R_0 & = & \frac{1}{2} \sum_{a,m} ( c^\dagger_{pam}c_{pam}+ c^\dagger_{nam}c_{nam}-1)
\end{eqnarray}
each couple $(P_-,P_0,P_+)$, $(N_-,N_0,N_+)$, $(R_-,R_0,R_+)$ generating an
su(2) algebra. Moreover, if $(T_-,T_0,T_+)$ are the isospin operators:
\begin{equation}
T_+=\sum_{a,m} c^\dagger_{nam}c_{pam}~~,~~T_-=\sum_{a,m} c^\dagger_{pam}c_{nam}
\end{equation}
\begin{equation}
T_0= \frac{1}{2} \sum_{a,m} (c^\dagger_{nam}c_{nam}- c^\dagger_{pam}c_{pam})
\end{equation}
then $(P_\pm,N_\pm,R_\pm,T_\pm,P_0,N_0)$ generate the second-rank semisimple
Lie algebra o(5) \cite{7}. Choosing $R_0$ and $T_0$ as basis elements in the
Cartan subalgebra, the root diagram can be pictured as in figure 1 \cite{8}.

\begin{figure}
\begin{picture}(100,215)(-150,-20)
\put(-100,100){\vector(1,0){200}}
\put(0,0){\vector(0,1){200}}
\put(95,80){$T_0$}
\put(5,190){$R_0$}
\put(50,85){$\frac{1}{\sqrt{6}}$}
\put(-50,85){$- \frac{1}{\sqrt{6}}$}
\put(-50,110){$T_-$}
\put(50,110){$T_+$}
\put(-15,150){$\frac{1}{\sqrt{6}}$}
\put(-25,50){$- \frac{1}{\sqrt{6}}$}
\put(50,155){$N_+$}
\put(5,155){$R_+$}
\put(-50,155){$P_+$}
\put(-50,40){$N_-$}
\put(5,40){$R_-$}
\put(50,40){$P_-$}
\put(0,100){\vector(1,0){50}}
\put(0,100){\vector(1,1){50}}
\put(0,100){\vector(0,1){50}}
\put(0,100){\vector(-1,1){50}}
\put(0,100){\vector(-1,0){50}}
\put(0,100){\vector(-1,-1){50}}
\put(0,100){\vector(0,-1){50}}
\put(0,100){\vector(1,-1){50}}
\end{picture}

{\small Figure 1. Root diagram of o(5) algebra. }
\end{figure}

The three commuting operators ${\cal P}^\dagger_{-1} = P_+$,
${\cal P}^\dagger_0 = R_+/ \sqrt{2}$, ${\cal P}^\dagger_1 = N_+$ are the components of
an isospin vector, and were used in \cite{9,10} to construct the
quadrupling four-particle interaction $H_Q= - G_Q Q_0^\dagger Q_0/4$, with
$Q^\dagger_0 = 2 \sum_{\mu=-1}^1 (-1)^\mu {\cal P}^\dagger_\mu
{\cal P}^\dagger_{- \mu}$. This interaction does not lead to the expected
gauge force for the superfluid systems, and consequently in the present work
a different choice will be made. Coupling the isovector ${\cal P}^\dagger_\mu$
with its Hermitian conjugate ${\cal P}_\mu$ to an isospin quadrupole $Q_{2 \mu}$,
an isoscalar four-particle interaction, separable in the two particles-two
holes channel will be defined by
\begin{equation}
H_4 = - G_4 \sum_{\mu=-2}^2 (-1)^\mu Q_{2 \mu}Q_{2 - \mu} =
- G_4 \sum_{\mu=-2}^2 Q_{2 \mu}^\dagger Q_{2 \mu}~~. \label{4pi}
\end{equation}
With this term, the total microscopic Hamiltonian, including the usual
pairing interaction, becomes
\begin{equation}
H=H_0+H_4
\end{equation}
\begin{equation}
H_0= \sum_{a,m} ( \epsilon_{pa} c^\dagger_{pam}c_{pam}+ \epsilon_{na}
c^\dagger_{nam}c_{nam}) - G_p P_+P_--G_nN_+N_- ~~.
\end{equation}
The ground state $\vert g \rangle$ of $H$ includes the 4-particle correlations
determined by $H_4$ and the transition to the gauge-restoring mean field is
expected when the sum rule:
\begin{equation}
\sum_{\mu=-2}^2 \langle g \vert Q_{2 \mu}^\dagger Q_{2 \mu} \vert g \rangle =
\sum_{\mu=-2}^2  \sum_{\vert n \rangle} \vert \langle n \vert
Q_{2 \mu} \vert g \rangle \vert^2
\end{equation}
is exhausted by $\vert g \rangle$:
\begin{equation}
\sum_{\mu=-2}^2 \langle g \vert Q_{2 \mu}^\dagger Q_{2 \mu} \vert g \rangle
\approx  \sum_{\mu=-2}^2 \vert q_\mu \vert^2~~,
\end{equation}
\begin{equation}
q_\mu = \langle g \vert Q_{2 \mu} \vert g \rangle
\end{equation}
and $q_{\pm 2} \ne 0$. This highly correlated ground state can be approximated
by the eigenfunction $\vert g_\omega \rangle$ of the linearised Hamiltonian $H_L$:
\begin{equation}
H_L = H_0 - G_4 \sum_{\mu=-2}^2 (q_\mu Q^\dagger_\mu + q_\mu^* Q_\mu) -
\omega T_0
\end{equation}
\begin{equation}
H_L \vert g_\omega \rangle = E_\omega \vert g_\omega \rangle
\end{equation}
where $q_\mu$ are self-consistently determined by:
\begin{equation}
q_\mu = \langle g_\omega \vert Q_{2 \mu} \vert g_\omega \rangle ~~.
\end{equation}
The "cranking" term $- \omega T_0$ was introduced for an approximate projection
of the proton and neutron numbers $Z,N$, because even if $H_L$ commutes with
$\hat{A}= \hat{N}_p+ \hat{N}_n$, it does not commute with $T_0$. So, the
parameter $\omega$ is fixed by the constraint:
\begin{equation}
\langle g_\omega \vert T_0 \vert g_\omega \rangle = \frac{N-Z}{2}~~.
\end{equation}
To obtain the function $\vert g_\omega \rangle$ further approximations will be
made, neglecting the proton-neutron pairing interaction and retaining in
$\vert g_\omega \rangle$ only the correlations between pairs of protons and
pairs of neutrons. Consequently, $H_L$ becomes:
\begin{equation}
H_L= \sum_{\tau ,a,m}  \epsilon_{ \tau a} c^\dagger_{\tau am}c_{\tau am}
 - \tilde{G}_p P_+P_- - \tilde{G}_nN_+N_-  + H_g - \omega T_0 ~~, \label{HL}
\end{equation}
where
\begin{equation}
\tilde{G}_\tau = G_\tau + \frac{2}{ \sqrt{6}} G_4 q_0
\end{equation}
and
$H_g = -2 G_4 (q_2^* N_+P_-+q_2N_-P_+)$, $q_2= \langle g_\omega \vert
N_+ P_- \vert g_\omega \rangle$ is the gauge restoring interaction. This
Hamiltonian will be restricted to one degenerate $j=3/2$ level. The interest
for this case is justified by the relevance of the O(5) symmetry for the
classification of the experimental data. So, the low-lying $0^+$ states of
the nuclei filling $j=3/2$ shells, as 1p$_{3/2}$, 1d$_{3/2}$ and 2p$_{3/2}$
can be assigned to the weight vectors of the 14-dimensional irreducible
representation of the o(5) algebra. In these multiplets both ground and
excited states are included, and figure 2 presents their relative energies with
Coulomb corrections\footnote{The experimental binding and excitation energies
have been taken from A. H. Wapstra and K. Bos, {\it The 1977 atomic mass
evaluation}, Atomic Data and Nuclear Data Tables {\bf 19} 177 (1977), and
P. M. Endt and C. van d\`er Leun, {\it Energy levels of A=21-44 nuclei},
Nucl. Phys. A {\bf 214} 1 (1973). When I obtained figure 2 (after the 
less symmetric diagram for the light nuclei filling the 1p$_{3/2}$ shell), I 
wondered if there was no previous attempt to make such a test of the O(5) 
symmetry. After a thorough search in literature for several weeks I found 
a practically unknown paper, by P. Camiz and U. Catani \cite{11}, in which 
the O(5) symmetry was used to classify nuclear states for a wide range of 
mass numbers. However, I could not find any indication that an effective 
proton-neutron interaction like $H_g$ of (\ref{HL}) has ever been used before.}.
Considering the same pairing constant $\tilde{G}$ and single-particle energy
$\epsilon$ for both protons and neutrons, the restricted Hamiltonian is
\begin{equation}
H_L= 2 \Omega \epsilon + ( 2 \epsilon + \omega) P_0 + (2 \epsilon - \omega ) N_0
\end{equation}
$$
 - \tilde{G} (P_+P_- +N_+N_-)  -2 G_4 (q_2^* N_+P_-+q_2N_-P_+) ~~,~~
 \Omega = j + \frac{1}{2}~~.
$$
This Hamiltonian commutes with the operator $\hat{A}$ and with the squares of
the proton and neutron quasispin operators $\vec{P}^2$, $\vec{N}^2$, so that
its eigenfunctions are labelled by the eigenvalues $E$, $A_v=$ number of the
valence particles, $p$ and $n$:
\begin{eqnarray}
H_L \vert pn E A_v \rangle_\omega & = & E_\omega \vert pnEA_v \rangle_\omega \\
\hat{A} \vert pn E A_v \rangle_\omega & = & A_v \vert pnEA_v \rangle_\omega \\
\vec{P}^2 \vert pn E A_v \rangle_\omega & = & p(p+1) \vert pnEA_v \rangle_\omega \\
\vec{N}^2 \vert pn E A_v \rangle_\omega & = & n(n+1) \vert pnEA_v \rangle_\omega~~.
\end{eqnarray}
In addition, $\vert pn E A_v \rangle_\omega$ depends on $\omega$ whose value
is fixed by the constraint (19):
\begin{equation}
~_\omega\langle pnEA_v \vert T_0 \vert pnEA_v \rangle_\omega =
\frac{N_v-Z_v}{2}~~. \label{om1}
\end{equation}

\begin{figure}
\begin{picture}(100,230)(-170,0)

\put(-150,0){\vector(0,1){220}}
\put(-159,0){0~\line(1,0){5}}
\put(-164,40){20 \line(1,0){5}}
\put(-164,80){40 \line(1,0){5}}
\put(-164,120){60 \line(1,0){5}}
\put(-164,160){80 \line(1,0){5}}
\put(-169.5,200){100 \line(1,0){5}}
\put(-140,215){E (MeV)}

\put(0,205){\small $0^+$;0}
\put(-5,196.194){\line(1,0){30}}
\put(0,185){\small $^{32}$S}

\put(80,163){\small $0^+$;1}
\put(80,155.098){\line(1,0){30} }
\put(80,143){\small $^{34}$Ar}

\put(0,163){\small $0^+$;1}
\put(0,156.08){\line(1,0){30}}
\put(0,143){\small $^{34}$Cl}

\put(-80,163){\small $0^+$;1}
\put(-80,155.09){\line(1,0){30}}
\put(-80,143){\small $^{34}$S}

\put(140,125){\small $0^+$;2}
\put(140,120.97){\line(1,0){30}}
\put(140,110){\small $^{36}$Ca}

\put(80,125){\small $0^+$;2 (?)}
\put(80,120){- - - - - }
\put(80,110){\small $^{36}$K}

\put(0,125){\small $0^+$;2 (10.858)}
\put(0,120.82){- - - - -}
\put(0,105){\small $0^+$;0}
\put(0,99.108){\line(1,0){30}}
\put(0,90){\small $^{36}$Ar}

\put(-80,128){\small $0^+$;2 (4.323)}
\put(-80,122.26){- - - - -}
\put(-80,110){\small $^{36}$Cl}

\put(-140,125){\small $0^+$;2}
\put(-140,122.32){\line(1,0){30}}
\put(-140,110){\small $^{36}$S}

\put(80,65){\small $0^+$;1}
\put(80,57.86){\line(1,0){30} }
\put(80,45){\small $^{38}$Ca}

\put(0,65){\small $0^+$;1(0.13)}
\put(0,57.834){- - - - -}
\put(0,45){\small $^{38}$K}

\put(-80,65){\small $0^+$;1}
\put(-80,57.854){\line(1,0){30}}
\put(-80,45){\small $^{38}$Ar}

\put(0,5){\small $0^+$;0}
\put(0,0){\line(1,0){30}}
\put(0,-10){\small $^{40}$Ca}

\end{picture}
\vskip.5cm
{\small Figure 2. The o(5) multiplet of J$^\pi=0^+$ ground (solid) and excited
(dash) states for the nuclei filling the 1d$_{3/2}$ valence level. Whenever
available, the excitation energy with respect to the ground state ($(2^+;1)$
for $^{36}$Cl, $^{36}$K and $(3^+;0)$ for  $^{38}$K) is given in round brackets
(MeV).
}
\end{figure}

\noindent
So, the function $\vert pnEA_v \rangle_\omega$ is associated with a given nucleus
from the isobaric family determined by $A_v$. The diagram in figure 2
contains five such families, having in the increasing order of the energies:
$A_v=$ 8, 6, 4, 2, 0. From these, the correlation between the proton and neutron
pairs is effective only for $p=n=\Omega/2=1$, $A_v = 2 \Omega =4$, and this case
will be discussed in the following. Denoting by $\vert p, p_0 \rangle$ and
$\vert n, n_0 \rangle$ the seniority eigenfunctions of $\vec{P}^2$, $P_0$
and $\vec{N}^2$, $N_0$, respectively, the function $\vert pnEA_v \rangle_\omega$ can
be expressed as:
\begin{equation}
\vert pnEA_v \rangle_\omega = \sum_{p_0+n_0=A_v/2 - \Omega} a_{p_0} \vert p,p_0 \rangle
\vert n,n_0 \rangle~~. \label{cwf}
\end{equation}
When $p=n=1$, $A_v=4$, $\Omega = 2$, the coefficients $a_{-1}$, $a_0$, $a_1$,
are easily found from the equation
$H_L \vert 11 E 4 \rangle_\omega = E_\omega \vert 11E4 \rangle_\omega$, and
(\ref{cwf}) becomes
\begin{equation}
\vert 11E4 \rangle_\omega = \frac{1}{\sqrt{1+8 \delta_\omega^2 \frac{ x^2 +
4 \omega^2}{(x^2 - 4 \omega^2)^2}}} ( \frac{ 2 \delta_\omega}{x+ 2 \omega} \vert 1,-1 \rangle
\vert 1,1 \rangle  \label{ev}
\end{equation}
$$
- \vert 1,0 \rangle \vert 1,0 \rangle
+\frac{ 2 \delta_\omega}{x- 2 \omega} \vert 1,1 \rangle \vert 1,-1 \rangle)~~.
$$
Here $x=2 \tilde{G} +E_\omega$ and $\delta_\omega = 2G_4q_2$, with $q_2$ chosen to
be real, self-consistently determined by
\begin{equation}
q_2=  ~_\omega\langle 11E4 \vert N_+P_- \vert 11E4 \rangle_\omega  
= \frac{8x \delta_\omega}{(4 \omega^2-x^2)(1+8 \delta_\omega^2 \frac{ x^2 +
4 \omega^2}{(x^2 - 4 \omega^2)^2}  )}~~.
\end{equation}
The $\omega$-dependent energy $E_\omega$ is analitically obtained from the
third-order equation
\begin{equation}
x^3+ 2 \tilde{G} x^2 - 4( 2 \delta_\omega^2 + \omega^2) x - 8 \tilde{G}
\omega^2 =0~~.
\end{equation}
Its solutions are
\begin{equation}
E_{g_\omega}= - \frac{8 \tilde{G}}{3} - \frac{4}{3} \cos \frac{\phi}{3}
\sqrt{\tilde{G}^2 + 6 \delta_\omega^2 + 3 \omega^2}~~,
\end{equation}
\begin{equation}
E_{k_\omega} = - \frac{8 \tilde{G}}{3} + \frac{2}{3}( \cos \frac{\phi}{3} + (-1)^k
\sqrt{3} \sin \frac{\phi}{3} ) \sqrt{\tilde{G}^2 + 6 \delta_\omega^2 + 3
\omega^2}~,~k=1,2 \label{ex}
\end{equation}
with  $\phi \in [0, \pi]$,
\begin{equation}
\tan \phi = \frac{\alpha}{\gamma}~~,
\end{equation}
\begin{equation}
\alpha = \sqrt{ \frac{64}{27} (\frac{\tilde{G}^2}{3}+2 \delta_\omega^2 + \omega^2)^3
-\gamma^2 }                       ~~,
\end{equation}
\begin{equation}
\gamma = \frac{8 \tilde{G}}{3} ( \frac{\tilde{G}^2}{9} + \delta_\omega^2 -
\omega^2 )      ~~,
\end{equation}
the system having two excited states. At $\omega =0$, the lower solution
$E_{g_0}$, corresponding to the ground state $\vert g_0 \rangle = \vert 11E_{g_0}
4 \rangle_0$, is\footnote{For the first excited state Eq. (\ref{ex}) yields
$E_{1_0}=-2 \tilde{G}$, while the corresponding state vector (\ref{ev}) is
$\vert 11 E_{1_0} 4 \rangle_0 = (\vert 1,-1 \rangle \vert 1,1 \rangle  - \vert 1,1 \rangle
\vert 1,-1 \rangle) / \sqrt{2}$.}
\begin{equation}
E_{g_0}=-3 \tilde{G} - \sqrt{\tilde{G}^2+8 \delta_0^2}~~.
\end{equation}
Using this value, the mean-field parameter $\delta_0 = 2 G_4 \langle g_0 \vert
N_+ P_- \vert g_0 \rangle $ can be easily calculated, and one obtains
\begin{equation}
\delta_0 = \sqrt{8 G_4^2 - \tilde{G}^2/8}~~.
\end{equation}
Differentiating $E_\omega$ with respect to $\omega$, the mean value of $T_0$ can be found
\begin{equation}
~_\omega\langle 11E4 \vert T_0 \vert 11E4 \rangle_\omega =- \frac{dE_\omega}{d \omega}
\vert_{\delta_\omega=const}
\equiv J^\omega \cdot \omega~~, \label{om2}
\end{equation}
\begin{equation}
J^\omega = \frac{ 4 ( 2 \tilde{G}+x)^2}{8 \tilde{G} \delta_\omega^2 -x
(2 \tilde{G}+x)^2}~~.
\end{equation}
The factor $J^\omega$ can be interpreted as the moment of inertia associated with
the rotations around the Z axis in isospace. It is not defined as positive,
however for the ground state at $\omega =0$ ($N_v=Z_v$) it proves to be positive:
\begin{equation}
J^0_g = \frac{ 32 \delta_0^2}{\sqrt{\tilde{G}^2+8 \delta_0^2}(\sqrt{\tilde{G}^2+8 \delta_0^2}
+ \tilde{G})^2}~~. \label{j0}
\end{equation}
If $\omega \ne 0$, but small, the energy $E_{g_\omega}$ can be approximated as:
\begin{equation}
E_{g_\omega} = E_{g_0} - \frac{ J^0_g \omega^2}{2}~~,
\end{equation}
and the ground state energy ${\cal E}_{g_\omega}$, defined by
${\cal E}_{g_\omega} = \langle g_\omega \vert H \vert g_\omega \rangle$,
becomes within the above approximations
\begin{equation}
{\cal E}_{g_\omega} = E_{g_0} + \frac{ J^0_g \omega^2}{2}~~. \label{eom}
\end{equation}
The parameter $\omega$ is calculated from the equations (\ref{om1}), (\ref{om2}), and for
small values $(N_v-Z_v)/2$ given to $\langle g_\omega \vert T_0 \vert g_\omega \rangle$, it 
will be
\begin{equation}
\omega = \frac{\langle g_\omega \vert T_0 \vert g_\omega \rangle}{J_g^0} =
\frac{N_v-Z_v}{2J_g^0}~~.
\end{equation}
Replacing $\omega$ from (\ref{eom}) by this value, the energies ${\cal E}_g(A_v;N,Z)$
of the even-even nuclei from the isobaric family having $A_v=4$ can be expressed as 
\begin{equation}
{\cal E}_g(4;N,Z) = E_{g_0} + \frac{(N-Z)^2}{8J^0_g}~~. \label{eom4}
\end{equation}
Using these results, some numerical estimates of the constants $G_4$ and
$\delta_0$ for the level 1d$_{3/2}$ will be made in the following. In this case,
the state $\vert 11E_{g_0} 4 \rangle_0$ having $N_v=Z_v=2$ corresponds to the
isosinglet ground state of $~^{36}$Ar, while $\vert 11E_g 4 \rangle_\omega$,
$\omega =(N_v-Z_v)/2J_g^0$, for  $N_v=4$, $Z_v=0$ and $N_v=0$, $Z_v=4$
correspond to the $T=2$ ground states of $^{36}$S and $^{36}$Ca, respectively.
As for the members of the same isomultiplet, the ground state energies
${\cal E}_g(4;4,0)$ and  ${\cal E}_g(4;0,4)$ of $^{36}$S and $^{36}$Ca are the
same, higher than the ground-state energy ${\cal E}_g(4;0,0)$ of $^{36}$Ar.
The difference ${\cal E}_g(4;4,0)-{\cal E}_g(4;0,0)$ is just the excitation
energy $E_{0^+;2}$ of the state $({\rm J}^\pi;T)=(0^+;2)$ of $^{36}$Ar at 10.858 MeV.
So, from (\ref{eom4}) a first estimate of $J^0_g$ can be obtained:
\begin{equation}
\frac{1}{J^0_g} = \frac{E_{0^+;2}}{2}~~.
\end{equation}
Now, the constant $G_4$ can be calculated from the second-order equation given
by (\ref{j0}). Following \cite{9} in taking $\tilde{G}=0.6$ MeV, two solutions
for $G_4$ are found: $G_4^{(1)}=2.56$ MeV, $G_4^{(2)}=0.08$ MeV. The solution $G_4^{(2)}$ is
closer to the value of the quadrupling interaction constant $G_Q=0.09$ MeV
\cite{9}, suggesting that $G_4=G_4^{(2)}$ is the correct choice, if both terms,
$H_Q$ and $H_4$ come out from the same basic four-particle interaction. For
this choice the coupling constant of the phenomenological gauge restoring
interaction in $^{36}$Ar has the value $\delta_0 = 0.07$ MeV.

\begin{figure}
\begin{picture}(100,215)(0,0)
\end{picture}
\vskip.2cm
{\small Figure 3. A comparison between the experimental energies $E_{0^+;2}^A$
(* symbols), and $W_{sym}(A,4)$ (solid line).}
\end{figure}

In the case of other $j=3/2$ valence levels the numerical values of the
parameters $\tilde{G}$, $\delta_0$, must be changed because they are dependent
on the mass number $A$ of the $T=0$ and $T=2$ isobaric isomultiplets. 
The average will be used for the pairing strength $\tilde{G}=21/A$ MeV of the proton and
neutron constants $\tilde{G}_p=23/A$ MeV,  $\tilde{G}_n=19/A$ MeV. The $A$-dependence
of the mean-field parameter $\delta_0$ is unknown, but it will be obtained from the 
expression of the isorotational energy. The term $(N-Z)^2/8 J^0_g$ appearing in (\ref{eom4}) 
represents the contribution of the nucleons lying on the same degenerate
$j=3/2$ level to the symmetry energy from the Weisz\"acker mass formula,
$W_{sym}(A, N-Z)=k_w (N-Z)^2$, $k_w=28.1/A$ MeV. This formula describes the
average symmetry energy of the nuclei, and in particular, for $N-Z=4$ it is
expected to approximate the energy $E^A_{0^+;2}$ in the light nuclei having
$N=Z=A/2$. A comparison with the experimental data shows that $W_{sym}(A,4)$
is systematically greater than $E^A_{0^+;2}$ (figure 3), but except for the mass region $A \le 28$
the relative difference   $W_{sym}(A,4)/E^A_{0^+;2}-1$ decreases with $A$,
becoming 1.5 \% for $^{52}$Fe. Consequently, for $A \ge 28$, $W_{sym}(A,4)$ reproduces
quite well the value of $E^A_{0^+;2}$, proving in particular that $J^0_g$ is proportional
to $A$. As $\tilde{G} \sim A^{-1}$, the formula (\ref{j0}) also implies an average
$A^{-1}$ dependence for $\delta_0$, which in the range $A \ge 28$ is determined by its value
at $A=36$ to be $\delta_0 = 2.7/A$ MeV. \\[.3cm]
{\it 3. The Fermi levels oscillations in superfluid nuclei} \\[.3cm] \indent
The ground state of the superfluid systems  contains correlations between
the same kind of particles, and is well approximated by the BCS function. For
a single proton level the pairing Hamiltonian $H_p$ and the BCS function become:
\begin{equation}
H_p= \epsilon \hat{N}_p - \tilde{G} P_+P_-
\end{equation}
\begin{equation}
\vert BCS \rangle_{ (\varphi,\lambda)} = e^{2zP_+-2z^*P_-} \vert 0 \rangle =
e^{-i \varphi \hat{N}_p/2} \vert BCS \rangle_{ (0, \lambda)}
\end{equation}
where $\lambda$ is the Fermi energy and $z=\rho e^{-i \varphi}$. Explicitly,
$\vert BCS \rangle_{ (0, \lambda)}$ is determined as the ground-state eigenfunction
for the Hamiltonian $H'=H_{pL}- \lambda \hat{N}_p$, containing the linearised
Hamiltonian $H_{pL}$:
\begin{equation}
H_{pL}= \epsilon \hat{N}_p - \Delta( P_+ +P_-)~~,
\end{equation}
\begin{equation}
\Delta = \tilde{G} \langle BCS_{ (0, \lambda)} \vert P_+ \vert BCS_{ (0, \lambda)}
\rangle = \frac{ \tilde{G} \Omega}{2} \sin 4 \rho~~.
\end{equation}
The term $- \lambda \hat{N}_p$ appearing here can be interpreted as a cranking
term restoring the symmetry broken by $H_{pL}$. To reach this interpretation,
one starts from the time-dependent Schr\"odinger equation $i \partial_t \vert
\Psi (t) \rangle = H_{pL} \vert \Psi (t) \rangle$ written for the Hamiltonian
$H_{pL}(t)= e^{-i \varphi (t)\hat{N}_p/2} H_{pL} e^{i \varphi (t) \hat{N}_p/2}$,
generated by a uniform gauge rotation with the angular velocity $\dot{\varphi}
= 2 \lambda$ from $H_{pL}$. Its solution $\vert \Psi (t) \rangle=
e^{-i \varphi (t) \hat{N}_p/2} \vert \Psi_0 (t) \rangle$, $\vert \Psi_0(t) \rangle=
e^{-i E'_g t} \vert BCS \rangle_{(0, \lambda)}$ contains the previous eigenfunction
of $H'$, $\vert BCS \rangle_{(0, \lambda)} $, whose eigenvalue $E_g'$ and
parameter $\rho$ are determined by 
\begin{equation}
H' \vert BCS \rangle_{(0, \lambda)} = E'_g \vert BCS \rangle_{(0, \lambda)}
\end{equation}
to be
\begin{equation}
E'_g = - E \Omega (1 - \frac{ \epsilon - \lambda}{E}) ~~,~~E=\sqrt{(\epsilon-
\lambda)^2 + \Delta^2}=\frac{\tilde{G} \Omega}{2}~~,
\end{equation}
\begin{equation}
\cos 4 \rho = 2 \frac{\epsilon - \lambda}{\tilde{G} \Omega}~~.
\end{equation}
Also, the energy function ${\cal E}$ and the mean number of particles ${\cal N}$ of the 
system are:
\begin{equation}
{\cal E} = \langle BCS_{(0, \lambda)} \vert H_p \vert BCS_{(0, \lambda)}
\rangle = \epsilon \Omega - \frac{ \epsilon^2 I_p}{2} - \frac{ \Omega^2}{2I_p}+
\frac{I_p \lambda^2}{2} + {\cal O}({\cal N} / \Omega),
\end{equation}
\begin{equation}
{\cal N} = \langle BCS_{(0, \lambda)} \vert \hat{N}_p \vert BCS_{(0, \lambda)} \rangle
= \Omega -  \epsilon I_p + \lambda I_p~~,
\end{equation}
where $I_p = 2/ \tilde{G}$ is the "moment of inertia" for the rotation
generated by $\hat{N}_p$, and ${\cal O}({\cal N}/ \Omega)=
-(\epsilon - \lambda - \tilde{G} \Omega /2)^2/ \tilde{G} \Omega$ is usually
neglected, assuming $\Omega$ large. A direct calculation shows that similar
formulae are obtained for the case of $2 \Omega$ levels distributed with a
constant density $\bar{ \rho} = 2 \Omega /(\epsilon_2 - \epsilon_1)$ between
the energies $\epsilon_2 > \epsilon_1$:
\begin{equation}
{\cal E}^c = \bar{\epsilon} \Omega - \frac{ \bar{\epsilon}^2 I_p^c}{2} - \frac{ \Omega^2}{2I_p^c}+
\frac{I_p^c \lambda^2}{2}~~,~~\bar{\epsilon} = \frac{\epsilon_1+\epsilon_2}{2}~~,
\end{equation}
\begin{equation}
{\cal N}^c = \Omega - \bar{ \epsilon} I_p^c + \lambda I_p^c~~,
\end{equation}
except for the moment of inertia which becomes $I_p^c = \bar{\rho} / \coth
(2/ \bar{\rho} \tilde{G})$ \cite{12}. \\ \indent
To investigate the effects of the interaction $-\delta_0 (N_+P_-+N_-P_+)$ for
a superfluid system, the microscopic Hamiltonian $H_L$ of (\ref{HL}) will be
treated semiclassically\footnote{This method is expected to yield the average properties 
of the collective excitation produced by the interaction term $-\delta_0 (N_+P_-+N_-P_+)$ 
when $\Omega$ is large, becoming increasingly accurate when $\delta_0 / \tilde{G}$ 
increases (M. Grigorescu, Can. J. Phys. {\bf 78} 119 (2000)).  }, using the variational 
method for the time-dependent trial function $\vert \Psi \rangle$. If $\{a_i, i=1,n \}$ are the real 
time-dependent parameters of $\vert \Psi \rangle$, then the variational equations:
\begin{equation}
\delta_{a_i} \int dt \langle \Psi \vert i \partial_t -H_L \vert \Psi \rangle =0 \label{tdvp}
\end{equation}
give $\vert \Psi \rangle$ as an approximation of the exact solution of the equation 
$i \partial_t \vert \Psi (t) \rangle =H_L \vert \Psi \rangle$. Explicitly
(\ref{tdvp}) becomes:
\begin{equation}
\sum_{j=1}^n B_{ij} \dot{a}_j = \frac{\partial {\cal H}}{\partial a_i}~~,~~
{\cal H}=\langle \Psi \vert H_L \vert \Psi \rangle~~, 
\end{equation}
\begin{equation}
B_{ij}=2 {\rm Im} \langle \partial_j \Psi \vert \partial_i \Psi \rangle ~~,~~
\partial_j \equiv \partial / \partial a_j~~,
\end{equation}
describing a classical dynamical system on the trial functions manifold \cite{13}.
Particularly, choosing $\vert \Psi \rangle$ as a product of two BCS functions,
one for protons and another for neutrons, $\vert \Psi \rangle = \vert BCS_p
\rangle \vert BCS_n \rangle$, this dynamical system becomes Hamiltonian. For
the one-level case, 
$$
\vert \Psi \rangle = e^{2z_pP_+-2z^*_pP_-} e^{2z_nN_+-2z^*_nN_-} \vert 0 \rangle~~,
~~z_\tau=\rho_\tau e^{-i \varphi_\tau}~~, \tau=p,n
$$
and the gauge angles together with the mean number of particles 
${\cal N}_\tau = \langle \Psi \vert \hat{N}_\tau \vert \Psi \rangle =2 \Omega
\sin^2 2 \rho_\tau$, give the canonical coordinates $\{ Q_\tau, P_\tau \}_{\tau=p,n}$:
\begin{equation}
Q_\tau=\varphi_\tau~~,~~P_\tau=\frac{{\cal N}_\tau}{2}~~.
\end{equation}
Their time evolution is determined by the Hamilton equations:
\begin{equation}
\dot{Q}_\tau =  \frac{\partial {\cal H}}{\partial P_\tau}~~,~~
\dot{P}_\tau = - \frac{\partial {\cal H}}{\partial Q_\tau}~~,
\end{equation}
with
\begin{equation}
{\cal H} = 2 (\epsilon - \frac{ \tilde{G} \Omega}{2})(P_p+P_n)+\tilde{G}( 1- 
\frac{1}{\Omega}) (P_p^2+P_n^2)
\end{equation}
$$
-2 \delta_0 \sqrt{P_pP_n(\Omega-P_p)(\Omega-P_n)} \cos(\varphi_p-\varphi_n)
$$
or equivalently, by the solutions $\{ \rho_\tau(t), \varphi_\tau(t) \}_{\tau=p,n}$
of the system
\begin{eqnarray}
\dot{\rho}_p & = & - \frac{\delta_0 \Omega}{4} \sin 4 \rho_n \sin (\varphi_p-\varphi_n) \\  
\dot{\rho}_n & = & \frac{\delta_0 \Omega}{4} \sin 4 \rho_p \sin (\varphi_p-\varphi_n) \nonumber \\
\dot{\varphi}_p & = & 2 \epsilon - \tilde{G} - \tilde{G}( \Omega -1) \cos 4 \rho_p - 
\delta_0 \Omega \cot 4 \rho_p \sin 4 \rho_n \cos(\varphi_p-\varphi_n) \nonumber \\
\dot{\varphi}_n & = & 2 \epsilon - \tilde{G}-\tilde{G} (\Omega-1)  \cos 4 \rho_n - 
\delta_0 \Omega \cot 4 \rho_n \sin 4 \rho_p \cos(\varphi_p-\varphi_n) ~~. \nonumber
\end{eqnarray}
If $\delta_0=0$, then $\dot{\rho}_p=0$, $\dot{\rho}_n=0$, and the number of
particles is fixed by the initial values $\rho_p(0)$, $\rho_n(0)$, without
additional constraints to the variational equation (\ref{tdvp}). When $\delta_0 \ne 0$ it
is convenient to use the "centre of mass" and "relative" canonical coordinates
$(Q,P)$ and $(\Phi, \wp)$, respectively, defined by:
\begin{eqnarray}
Q & = & \frac{Q_p+Q_n}{2}~~,~~ P  =  P_p+P_n  \\
\Phi & = & Q_p-Q_n ~~,~~ \wp  = \frac{P_p-P_n}{2}~~.
\end{eqnarray}
In these coordinates ${\cal H}$ becomes
\begin{equation}
{\cal H}(P, \wp,\Phi) = 2 (\epsilon - \frac{ \tilde{G} \Omega}{2})P+ 
\tilde{G} (1 - \frac{1}{\Omega}) ( \frac{P^2}{2}+2 \wp^2)
\end{equation}
$$
-2 \delta_0 \sqrt{(\frac{P^2}{4}-\wp^2)[(\Omega- \frac{P}{2})^2- \wp^2]} \cos \Phi
$$
It is appearing clearly that $P=A/2$, $A={\cal N}_p+{\cal N}_n$, is time-independent.
Moreover, the set $M=\{(Q,P,\Phi, \wp), \Phi=0, \wp=0 \}$ consists of closed
orbits, distinguished by $A$. For each $A=0,4,8,...,4 \Omega$ there is an orbit
in $M$ having $\Phi (t)=0$, $\wp (t)=0$, $P(t)=A/2$, and $Q(t)=Q_0+ 2 \lambda t$,
with $2 \lambda=2 [\epsilon - (\tilde{G}+\delta_0) \Omega/2] + [\tilde{G}
(1- \Omega^{-1})+\delta_0] A/2$. 
If $(\Phi, \wp)$ are not zero, but small, the orbits can be found by
expanding ${\cal H}(P, \wp, \Phi)$ around the point $(P,0,0)$. Within this
approximation ${\cal H}$ becomes:
\begin{equation}
{\cal H} = 2 (\epsilon - \frac{ \tilde{G} + \delta_0}{2} \Omega)P+
 [\tilde{G}( 1- \frac{1}{\Omega}) +\delta_0] \frac{P^2}{2}+ k \wp^2+\frac{C \Phi^2}{2} ~~,
\end{equation}
\begin{equation}
k=2\tilde{G}( 1- \frac{1}{\Omega})-2\delta_0-\frac{16 \delta_0 \Omega^2}{A(A-4\Omega)}~~,~~
C=2 \delta_0 (\frac{\Delta}{\tilde{G}})^2~~,
\end{equation}
and the equations describing the time evolution of the relative coordinates
$(\Phi, \wp)$ can be easily integrated to:
\begin{equation}
\Phi(t)=\Phi_0 \sin \omega_v t~~,~~ \wp (t)= \wp_0 \cos \omega_v t ~~,
\end{equation}
\begin{equation}
\wp_0 = \frac{\omega_v}{2k} \Phi_0~~,~~ \omega_v=\sqrt{2kC}~~.
\end{equation}
The interpretation of the derivatives $\dot{Q}_p/2$, $\dot{Q}_n/2$ as Fermi
energies associates the solution given above for the $N=Z$ ($\wp=0$) nuclei,
to small oscillations in opposite directions of the proton and neutron Fermi
levels. Pictorially, in the Dasso-Vitturi representation \cite{14}, they
correspond to relative angular oscillations of the proton and neutron
deformed densities in gauge space. Such oscillations, having an isovector
character, are allowed by the diffuseness of the Fermi surface for the
superfluid systems. Their quantisation \cite{15} leads to a harmonic oscillator
spectrum, whose first excited state has the energy $\omega_v$\footnote{When 
$A=4$, $\Omega =2$ this vibration mode correponds to the first excited state 
$\vert 11E_14 \rangle_0$, with the excitation energy $e_x= \tilde{G}+
(\tilde{G}^2+8 \delta_0^2)^{1/2}$. The difference $e_x-\omega_v$ is 
positive but decreases with $\delta_0$, and for $\delta_0 > 1.5 \tilde{G}$ it
falls below 10 \% of $e_x$.}. For numerical
estimates of this energy it is convenient to replace the constant $k$ obtained
above for a single level by the realistic value $k=16 k_w=449.6/A$ MeV, as given
by the symmetry energy. The constant $C$ will be calculated for $N=Z$ nuclei
with half-filled valence shells ($\sin 4 \rho =1$), taking $\delta_0=2.7/A$ MeV
as before, and $\Delta = \tilde{G} \Omega /2$. Here $\tilde{G}=21/A$ MeV, and
$\Omega \approx 0.5 (3 A/2)^{2/3}$ is half the Fermi level degeneracy for the
harmonic oscillator potential \cite{2}. Using these constants, the energy of the
isovector monopole vibration becomes $\omega_v = 22.8 A^{-1/3}$ MeV. \\[.3cm]
{\it 4. Conclusions } \\[.3cm] \indent
The gauge-restoring interaction $-\delta_0 (N_+P_-+N_-P_+)$ introduced in order
to fix the same proton and neutron Fermi levels in superfluid nuclei was proved
to be strongly connected with the symmetry energy and with the four-particle
interaction. For the non-superfluid nuclei, its non-commutation with $T_0$ leads
to a kinetic energy term corresponding to collective rotations around the
Z-axis in isospace (isorotations). In contrast to the purely kinematic symmetry energy
determined by the Pauli principle in the Fermi gas model \cite{16}, the
isorotational symmetry energy occurs dynamically from the proton-neutron
correlations. This result is supported by the experimental data on light nuclei,
which show large mass differences $E_{0^+;2}$ between the even-even
members of the $T=0$ and $T=2$ isomultiplets having the valence nucleons on the
same degenerate level.
An energy $E_{0^+;2}$ close to the value predicted by the symmetry energy
term from the Weisz\"acker mass formula in its simplest form, was taken as an
indication of the isorotational origin of the $T=2$ multiplet. This was the case
for the even-even isobars with $A \ge 28$, and the energy $E_{0^+;2}$
corresponding to $A=36$ was used to estimate the constant $\delta_0$ as
$\delta_0 = 2.7/A$ MeV. Because
the microscopic proton-neutron interaction is number conserving, $\delta_0$
was considered to be a mean-field parameter. Such a mean field corresponds
to the four-particle interaction expressed in (\ref{4pi}), whose strength $G_4$
gives $\delta_0$ self-consistently. Estimating $G_4$ for $^{36}$Ar, a value close
to the quadrupling constant $G_Q$ previously used in \cite{9} was
obtained. This result shows that the isospin quadrupole interaction (\ref{4pi}) may
represent an alternative to quadrupling for the charge-independent treatment
of the nuclear correlations. \\ \indent
In the superfluid nuclei the gauge-restoring interaction leads to isovector
oscillations of the Fermi levels, superposed over uniform isorotations.
Their energy $\omega_v=22.8 A^{-1/3}$ MeV was estimated for $A \ge 28$ nuclei with
half-filled valence shells, considering the same constant $\delta_0$ as for the
$j=3/2$ shell. Due to the low value of $\omega_v$ the Fermi levels
oscillations are different from the $170  A^{-1/3}$ MeV resonances presented
in \cite{3,4,5}, but can appear as low-lying IVMR in high resolution charge
exchange or electron scattering experiments. Their occurrence is
expected for the superfluid nuclei on the $\beta$ stability line, the further
experimental investigation representing a test for the gauge angles dynamics
presented above.

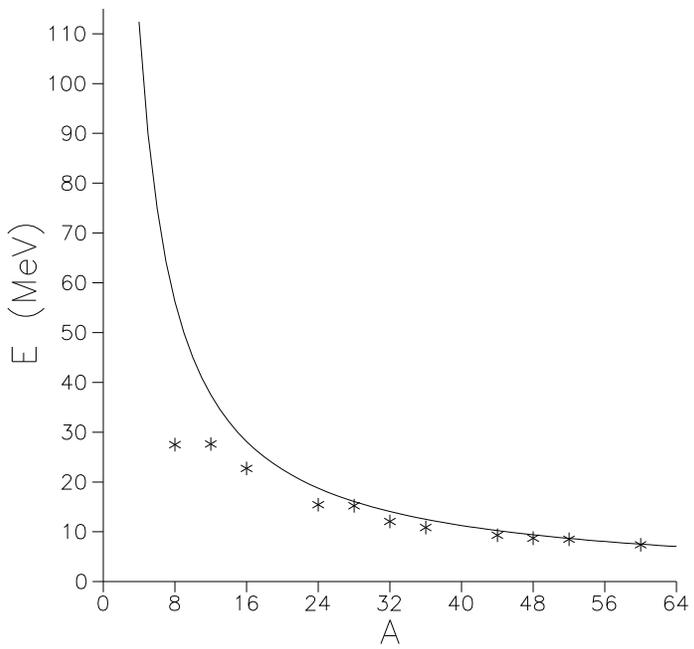
\end{document}